# Multiple timescale contact charging

Troy Shinbrot[1,2*], Behrooz Ferdowsi[1,3], Sankaran Sundaresan[2], Nuno A.M. Araujo[4,5]

[1] Department of Biomedical Engineering, Rutgers University, Piscataway, NJ 08816, USA
[2] Department of Chemical and Biological Engineering, Princeton University, Princeton, NJ 08544, USA
[3] Department of Geoscience, Princeton University, Princeton, NJ 08544, USA
[4] Departamento de Física, Faculdade de Ciências, Universidade de Lisboa, 1749-016 Lisboa, Portugal
[5] Centro de Física Teórica e Computacional, Universidade de Lisboa, 1749-016 Lisboa, Portugal
* shinbrot@rutgers.edu

**Abstract:**

Contact charging between insulators is one of the most basic, yet least well understood, of physical processes. For example we have no clear theory for how insulators recruit enough charge carriers to deposit charge but not enough to discharge. In this letter we note that charging and discharging kinetics may be distinct, and from this observation we develop a mathematical model. The model surprisingly predicts that charging can decrease as contact frequency increases: we confirm this prediction experimentally and propose future steps.

The most basic observation of static electrification, dating to William Gilbert's work in the 16[th] century[1], is that the materials that most readily acquire contact charge are insulators. We all know this: we get a shock when walking with rubber soles on a nylon carpet, on dry days when there are no free charge carriers to be found. Based on experiments with as many materials as he could lay his hands on, Gilbert termed materials that develop static charge "electrics:" silk, amber, glass and the like. These are what we today call insulators: materials that have no free charge carriers.

This is undeniably very odd: if charges are bound, they should not transfer between materials, and if they are free they should conduct charges away. Insulators manage to mobilize sufficient free charges to transfer charge, but not enough to discharge. In this letter we focus on a particular, and revealing, feature of this oddity.

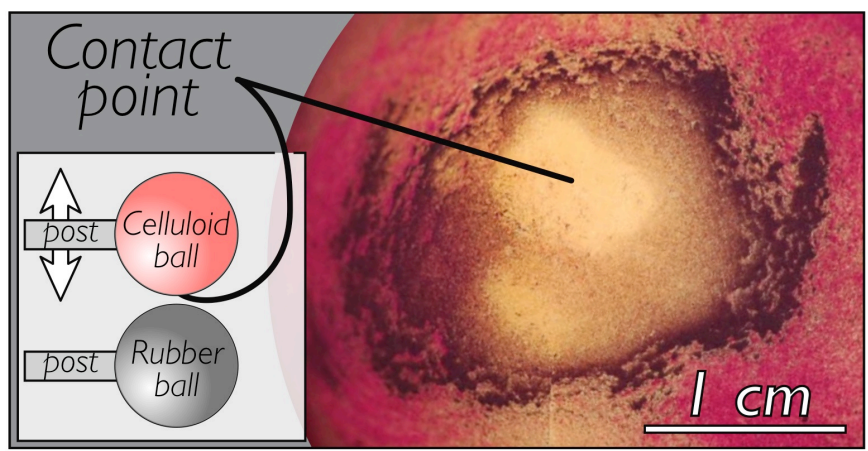

*Figure 1 – Example of contact charge pattern, from [3]. Inset summarizes experiment, in which a celluloid ball mounted on a wooden post is contacted with a synthetic rubber ball on a wooden post. Main plot shows charge patterns made visible by exposing the celluloid ball to a cloud of bipolar toner[2]: the red (black) toner sticks to negatively (positively) charged areas. Note that charge remains on the ball centimeters away from the contact point, and that point retains minimal charge. This suggests different spatial and temporal scales for charging and discharging.*

Consider Fig. 1, described in detail in Ref. [3]. The inset depicts an experiment in which a celluloid (ping-pong) ball attached to an insulating post is vibrated so as to contact a synthetic rubber ball beneath, also attached to an insulating post. No conductors are within 10 cm of the experiment, and after vibrating at 10 Hz for 30 seconds, the ping-pong ball is exposed to a cloud of bipolar toner[2], which reveals the charge patterns shown in the main plot of Fig. 1. The red toner is positively charged, and so it sticks to negative regions on the ball, and the black toner is negative and so it highlights positive charges. Prior to contact, very little toner sticks to the ball (see Ref. [3]).

In the present letter, we seek to understand one aspect of the mechanism that produces charging of insulators in this way. Fig. 1 provides two significant clues to this mechanism. First, following repeated contact this insulating ball has indeed liberated enough charge carriers to electrify its surface, but the process has also liberated enough charge carriers to *neutralize* a sizeable region near the point of contact. So both charging and discharging appear to be at work. Second, the ultimate charge transported exhibits multiple spatial scales: the positive, black, halo is on the order of centimeters across, and the irregular boundary separating positive and negative regions is on the order of millimeters across. So apparently the charging process has transported charges a distance of centimeters, but that process cannot transport charges across a gap of millimeters to neutralize the final charge pattern.

We therefore propose that charging and discharging may involve different processes, and, critically, that the characteristic timescales of these processes may differ as well. For our purposes we are agnostic concerning details of the charge carriers involved, but we remark that it would not be unreasonable to speculate that surface-adsorbed ions[4] could transfer charge with low mobility in response to surface potential differences or applied fields[5], while bulk electrons raised to the conduction band by large electric fields could mediate discharges at high mobility.

This notion by itself is not sufficient to explain charging, because if discharging is much faster than charging, charges won't grow, whereas if charging is much faster than discharging, a single contact will produce charge separation, but multiple contacts will build no additional charge (by contrast, charge growth is observed experimentally[6,7,8]). Between these extremes, if charging is somewhat faster than discharging, we will show that charge growth can occur.

To construct a concrete model, we note that existing studies[3,9,10,11] have provided evidence that contact charging between particles may be mediated by two separate processes: polarization and neutralization. That is, an insulating particle can acquire an induced polarization due to the electric field of a neighboring particle – and when these particles come into contact, neutralization of part of that polarization can occur, leaving both particles with a residual charge.

Using this proposition as a starting point, we consider a simple, 1D, system, in which the magnitude of polarization, $p = |\vec{p}|$, of a particle can be written as:

$$\frac{dp}{dt} = r \cdot [\chi E - p], \qquad [1]$$

where $r$ is the rate of polarization due to surface ion mobility, $\chi$ is the electric susceptibility, and $E = |\vec{E}|$ is the electric field due to contact potentials or external charges. Here a particle will approach polarization $p = \chi E$ with exponential rate $r$, and to allow charging and discharging to proceed at different rates, we set:

$$r = \alpha \quad \text{if } p < \chi E$$
$$r = \beta \quad \text{otherwise.} \qquad [2]$$

Thus $\alpha$ defines the rate of growth of polarization, and $\beta$ defines the rate of its relaxation.

In other words, we consider here a scalar polarization that grows at rate $\alpha$ when the applied field is large enough to drive further polarization, and decays at rate $\beta$ when the applied field is smaller than that. We seek a solution in the case where two mirror-symmetric particles make repeated collisions with one another as sketched in the inset to Fig. 2(a). We emphasize that Eq's [1]-[2] represent a lumped element approach intended to describe the spatially averaged polarization of contact electrified particles. The charging mechanisms leading to the pattern shown in Fig. 1 are manifestly complex, as the figure shows. Our goal here is only to analyze the effects of multiple timescales on mean particle polarization, leaving detailed spatial as well as temporal analysis to future studies.

With this caveat in mind, as polarized particles make and break contact, the field on each will periodically grow and diminish – and so the rate $r$ should periodically flip between values $\alpha$ and $\beta$. In the simple case where the polarizations on each particle are identical and where particles collide with frequency $f$, the electric field on either particle will grow and diminish as the particles vibrate toward and away from one another. At its simplest, the field can be written:

$$E \simeq [A + B sin(ft)] \cdot p, \qquad [3]$$

where $A$ and $B$ are leading order Fourier coefficients. Here $A$ defines the mean field due to an oscillating polarized particle, and $B$ defines the periodic variation in field due to periodic oscillations. Inserting Eq. [3] into Eq. [1] completes the description of the problem depicted in Fig. 2(a).

The essential kinetics of Eq. [1] are of course exponential growth or relaxation. Yet any practical experiment will be bound by an ultimate limit, $p_b$ and its associated rate, $r_b$, due either to breakdown of the particle material or of its environment. Assuming that breakdown is independent of contact charging and discharging, we add a final term to Eq. [1] as follows:

$$\frac{dp}{dt} = r \cdot [\chi E - p] + r_b \cdot [p_b - p], \qquad [4]$$

where $r_b$ is a fixed rate if the polarization is greater than the breakdown limit, and is zero otherwise. So $r_b$ defines the rate of decay of polarization due to breakdown, and $p_b$ defines the critical polarization above which breakdown occurs.

We integrate Eq's [2]-[4] to produce time evolutions of $p$ shown in Fig. 2(b) for several values of vibration frequency, $f$. We include a time averaged plot in each case (solid curve) as well as the asymptotic value, ‹p› (broken line), which we define to be the mean of the polarization between times 800 and 1000. Parameters used are illustrative and are included in the caption to Fig. 2: crucially the charging rate $\alpha = 1$ is higher than the discharging rate $\beta = 0.5$, and the frequencies shown are in the range between the two values.

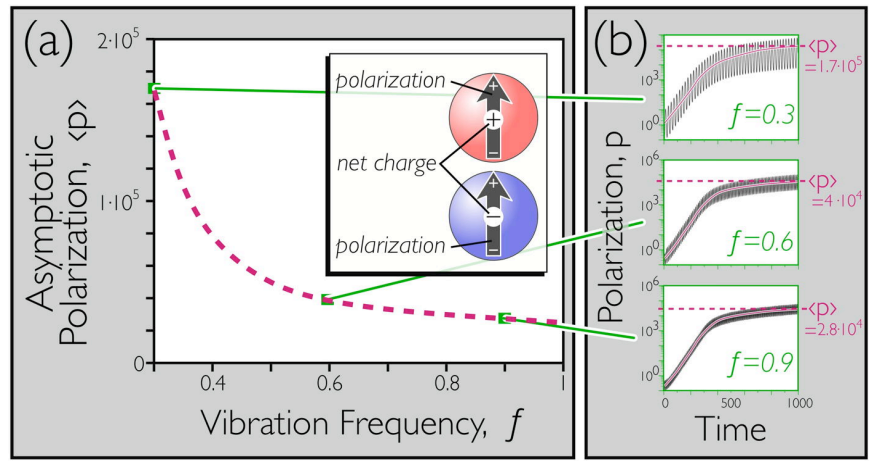

*Figure 2 – Simulations of repeated collisions between mirror-symmetric charged particles. (a) Inset: depiction of polarizations and charges on each particle: these generate an electric field on its partner, modeled in Eq's [1]-[2]. Here the polarizations are identical, and the net charges are equal and opposite, and we describe the field on either particle due to its mate in Eq. [3]. Main plot: integration of Eq's [2]-[4] using $A = 0.825$, $B = 1$, $\chi = 1$, $\alpha = 1$, $\beta = 0.5$, $p_b = 1000$, and when $p › p_b$, $r_b = 0.03$. (b) Exemplar growths in polarization on semilog axes for f=0.3, 0.6 and 0.9: notice that as the frequency, f, of particle collisions increases, the mean asymptotic polarization, ‹p›, decreases. Solutions shown are obtained using Mathematica's multistep Adam's method solver, NDSolve.*

Examination of the asymptotic polarization, ‹p›, reveals that it *decreases* as the frequency increases between the charging and the discharging rate. To make this more evident, in the main plot of Fig. 2(a), we display the dependence of ⟨$p$⟩ on frequency, $f$. Broadly speaking, the paradoxical decrease in polarization with increasing frequency occurs when the charging rate, $\alpha$, is higher than the discharging rate, $\beta$, and the driving frequency is of the same order as $\alpha$ and $\beta$.

To understand this heuristically, let's consider the polarization growth in several frequency ranges. When $f$ is much slower than $\alpha$ and $\beta$, a particle's polarization is quasi-static, and would in the absence of breakdown simply be slaved to the driving field. This means that without an external field[9], particle charges won't grow.

At the other extreme, when $f$ is much faster than $\alpha$ and $\beta$, oscillations defined in Eq. [3] become too fast for either charging or discharging kinetics to respond to. In this case, the mean field, $A$, can be expected to prevail. If $A ‹ p/\chi$, Eq. [2] tells us that the particle will discharge, whereas if $A › p/\chi$, Eq. [1] implies that the particle polarization will grow until it reaches the mean value $p = \chi A$. In which case again barring an external field, the particle will discharge. So for large $f$, the particle will either discharge (if $\chi$ and $A$ are small) or charge to the prescribed limit, $p = \chi A$ (i.e. the expected induced dipole moment).

Finally at intermediate values, for $f$ near $\alpha$ and $\beta$, the induced polarization will grow as the particles approach, but will seldom reach the threshold value $p = \chi E$ given by Eq. [2], and so discharging will be triggered less often than charging. Consequently the polarization will ratchet upward with time until it reaches a saturation value given by a balance between polarization growth and breakdown, as shown in Fig. 2(b). The faster the vibration frequency, $f$, however, the more often discharging will occur, and *so the ultimate polarization will decrease with $f$*.

Conveniently, the behavior shown in Fig. 2 implies that there is a simple test to assess the multiple timescale model: we can vibrate a bed of insulating grains and ask whether charging in fact decreases as vibration frequency increases. This would not be expected otherwise, for one would naively predict that, irrespective of what process produces charging, repeating the process more often should produce more charging.

We have therefore performed an experiment on a vibrated bed of 1 mm diameter glass beads in a polycarbonate box, shown in Fig. 3(a). We use a bed of particles rather than a single pair because using numerous particles allows us to average over multiple oscillating states (seen in Fig. 2(b)), and greater numbers of particles provide for greater growth in polarization than would be expected using only 2 particles[12]. The granular bed is ~ 1.5 cm deep at rest, and we suspend an electrostatic voltmeter probe (Trek, Model 347) above the bed, and vibrate the bed between 50 and 100 Hz, maintaining the shaker amplitude at 3±0.7g (measured with the accelerometer shown). We have also investigated a wider range of frequencies; we display a limited range here because the bed is uniformly fluidized and no Faraday patterns (which complicate results at lower frequencies) are seen. We carefully adjust the probe height at each frequency so that it is 14 cm above the apparent top of the vibrated bed surface (which again is nearly uniform at these frequencies), and we vibrate the bed until the probe records a steady voltage (about a minute). We show 27 frequencies, where each data point is an average over 3 replicates, and we perform the 81 trials in random order to eliminate systematic bias.

As shown in Fig. 3(b), *as the rate of vibration increases, the measured voltage does decrease*: here by about a factor of 3, strongly supporting the proposition that charging of the granular bed depends on a multiple timescale mechanism. To guide the eye we also copy and rescale a segment of the model plot from Fig. 2(a): agreement is far from definitive, but the model curve is consistent with the experimental data.

We remark that the voltage drops rapidly when vibration is stopped, which would not occur if the measurements were produced by a growth in net charge. This is also consistent with charge conservation in an insulating environment, and supports the analysis of dipole moment growth presented in Eq's [2]-[4]. The glass particles are, however, expected to tribocharge due to contact with the polycarbonate box: we speculate that this tribocharging may establish a field that initiates the polarization growth that we have defined. Tribocharging likely also contributes to the voltage measurements, however the *reduction* in voltage as frequency is increased is predicted by our model, but does not fit with current understanding of contact electrification[5].

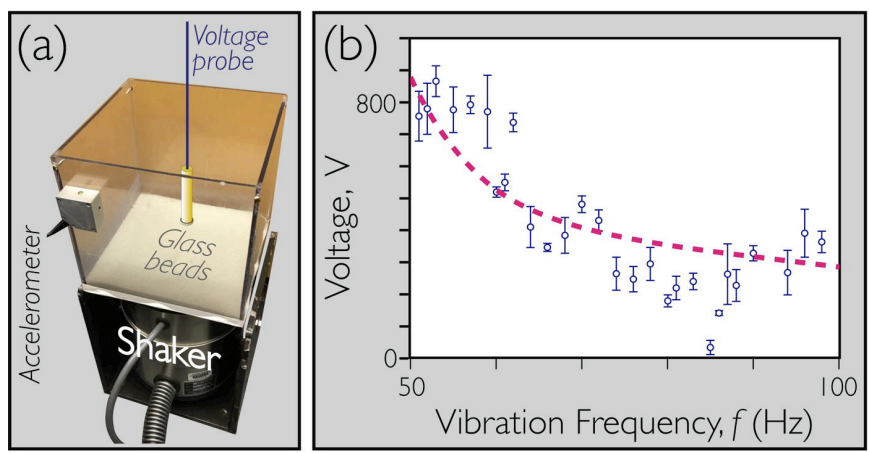

*Figure 3 – Experiment using vibrated bed to evaluate predicted decrease in charging with an increase in contact frequency. (a) Vibrated bed of glass beads; distance between probe and top of bed held fixed as described in text. (b) Plot of experimental measurements of voltage vs. vibration frequency confirming the predicted decrease in charge generation with an increase in frequency. Broken line is copied and rescaled from Fig. 2(a) between f = 0.4 and f = 1.*

Finally, it is important to stress that contact electrification has long been investigated[13] and is exceedingly complex[14]. Without question, contact charging variations depend both on material transport and on electronic densities of states at material interfaces[15], and these densities vary strongly with deformations[16] that occur during contact. Notwithstanding these complexities, we emphasize that the analysis of multiple timescale charging presented here represents the only model that predicts a decrease in charging with an increase in contact frequency.

In conclusion, we have developed a model for growth of contact-mediated polarization of insulating particles based on the proposition that charging and discharging obey different timescale kinetics. The model produces the counterintuitive prediction that when discharging kinetics are more rapid than charging kinetics, charging may decrease with increasing frequency of contacts. We confirm this paradoxical prediction in a vibrated bed experiment.

The model is highly simplified and provides considerable room for future extensions and analysis. For example, the model only considers the 1D problem of polarization of a single particle with its companion mirror particle. Prior work[11] has shown that in higher dimensions, colinear polarizations align, while non-colinear polarizations form anti-parallel pairs, so it is not obvious how vector polarizations in 2 or 3 dimensions will behave. Likewise in Fig. 1 we showed charge patterns on a single large particle, indicating that nontrivial spatial kinetics are also at work. How these spatial kinetics interact with the multiple timescale kinetics that we have proposed is both an intriguing and a difficult question, and deeper study is needed. Finally, the toner used to produce Fig. 1 only sticks to the outermost charges on the surface: it is not possible at this time to discriminate between net surface charge and a polarized double layer on the surface. The surface chemistry and surface-bulk interactions involved in both the charge kinetics and the ultimate patterning remain to be understood. We look forward to further developments in this very basic, yet unexpectedly rich, area.

## Acknowledgments:

We thank Jari Kolehmainen for helpful comments and analysis, and Brandon Jones and Pranav Saba for vital experimental assistance. We also acknowledge support from the NSF DMR award #1404792, CBET award #1804286, the Luso-American Development Foundation (FLAD), FLAD/NSF, Project #273/2016, and the Portuguese Foundation for Science and Technology (FCT) under Contract #UID/FIS/00618/2013.